\documentclass[12pt,preprint]{aastex}
\slugcomment{Submitted to The Astrophysical Journal Letters}

\shorttitle{Detection of HCO$^{+}$(5--4) line emission in APM~08279+5255 at z=3.9}

\shortauthors{Garc\'{\i}a-Burillo et al.}

\begin{document}


\title{A new probe of dense gas at high redshift: detection of HCO$^{+}$(5--4) line emission in APM~08279+5255 \footnote{Based on observations carried out with the IRAM Plateau de Bure Interferometer. IRAM is supported by INSU/CNRS (France), MPG (Germany) and IGN (Spain)}}

\author{S. Garc{\'{\i}}a-Burillo\altaffilmark{1}, J. Graci\'{a}-Carpio\altaffilmark{1}, M. Gu\'elin\altaffilmark{2}, R. Neri\altaffilmark{2}, 
P. Cox\altaffilmark{2}, P. Planesas\altaffilmark{1}, P.~M. Solomon\altaffilmark{3}, L.~J. Tacconi\altaffilmark{4}, P. A. Vanden Bout\altaffilmark{5}}
\altaffiltext{1}{Observatorio Astron\'{o}mico Nacional (OAN), C/ Alfonso XII 3, 28014 Madrid, Spain; s.gburillo@oan.es, j.gracia@oan.es,  p.planesas@oan.es}
\altaffiltext{2}{Institut de Radio Astronomie Millim\'etrique (IRAM), 300 Rue de la Piscine, Domaine Universitaire de Grenoble, F-38406 St. Martin d'H\`eres, France; guelin@iram.fr, neri@iram.fr, cox@iram.fr}
\altaffiltext{3}{Department of Physics and Astronomy, State University of New York at Stony Brook, Stony Brook, NY 11974; psolomon@astro.sunysb.edu}
\altaffiltext{4}{Max-Planck-Institut f\"ur extraterrestrische Physik, Postfach 1312, 85741 Garching, Germany; linda@mpe.mpg.de}
\altaffiltext{5}{National Radio Astronomy Observatory (NRAO), 520 Edgemont Road, Charlottesville, VA 22903; pvandenb@nrao.edu}



\begin{abstract}

We report the detection of HCO$^+$~(5--4) emission from the Broad
Absorption Line (BAL) quasar APM~08279+5255 at z=3.911 based on
observations conducted at the IRAM Plateau de Bure
interferometer. This represents the first detection of this molecular
ion at such a high redshift.  The inferred line luminosity,
uncorrected for lensing, is
$L'_{HCO^+}$=(3.5$\pm$0.6)$\times$10$^{10}$K\,km~s$^{-1}$\,pc$^2$. The
HCO$^+$ J=5--4 source position coincides within the errors with that
reported from previous HCN J=5--4 and high-J CO line observations of
this quasar.  The HCO$^+$ line profile central velocity and width are
consistent with those derived from HCN.  This result suggests that
HCO$^+$~(5--4) emission comes roughly from the same circumnuclear
region probed by HCN. However, the HCN~(5--4)/HCO$^+$~(5--4) intensity
ratio measured in APM~08279+5255 is significantly larger than that predicted 
by simple radiative transfer models, which
assume  collisional excitation and equal molecular abundances. This
could imply that the [HCN]/[HCO$^+$] abundance ratio is particularly
large in this source, or that the J=5 rotational levels are
predominantly excited by IR fluorescent radiation.

\end{abstract}

\keywords{galaxies: active --- galaxies: high-redshift --- galaxies: ISM --- galaxies: starburst --- ISM: molecules --- radio lines: galaxies}



\section{Introduction}

The quasar APM~08279+5255 at redshift $z$=3.91 is one of the most
luminous sources in the universe even after correcting for the high
lensing factor ($\sim$7) of its huge measured infrared luminosity
(L$_{IR}\sim$10$^{15}$L$_{\sun}$). Emission of high-J CO lines (J=9--8
and 4--3), mapped by \citet{Downes99} with the IRAM Plateau de Bure
Interferometer (PdBI), suggests the presence of a circumnuclear disk
of hot and dense molecular gas. The VLA detected also the emission of
the CO (1--0) and CO (2--1) lines in this source
\citep{Papadopoulos01, Lewis02}. The question of how much dense
molecular gas lies inside APM~08279+5255 was revisited by
\citet{Wagg05} who reported the detection of HCN~(5--4) emission in
this quasar using the IRAM PdBI.  The exceptionally strong intensity
of HCN~(5--4) with respect to all CO lines measured in APM~08279+5255
leaves room for different interpretations however. First, if the
excitation of all lines is mainly collisional, the high HCN/CO
luminosity ratios could stem from a large enhancement of the abundance
of HCN relative to CO or to a comparatively higher gas density for the
gas emitting in HCN lines. Alternatively, this could reflect
non-collisional excitation of high-J HCN lines through infrared
pumping around the active galactic nucleus (AGN).

Opposing theoretical scenarios can explain why the HCN/CO abundance
ratio may be anomalous in APM~08279+5255. HCN abundances could be
enhanced relative to other molecular species under the influence of
intense X-ray emission from the AGN \citep{Lepp96, Maloney96}. On the
other hand, selective oxygen depletion could decrease the abundance of
oxygen-bearing molecular species (e.g., CO) around AGNs
\citep{Sternberg94}. The use of different tracers that aim at
quantitatively probing the dense molecular gas in luminous and
ultraluminous infrared galaxies (LIRGs:L$_{IR}>$10$^{11}$L$_{\sun}$
and ULIRGs:L$_{IR}>$10$^{12}$L$_{\sun}$) is mandatory to elucidate
between the conflicting scenarios. To date, HCN has been widely used
to trace dense gas in galactic cores, nearby galaxies and
even ULIRGs. Of particular note, the luminosity of the HCN~(1--0)
line shows a remarkable correlation with L$_{IR}$
\citep[e.g.,][]{Solomon92, Gao04a, Gao04b, Wu05}.  However, recent
results of a combined HCN and HCO$^{+}$ survey of a sample of LIRGs
and ULIRGs which show a larger HCN/HCO$^+$ intensity ratio
towards ULIRGs with powerful AGNs, has been used to question
the validity of HCN as a quantitative tracer of dense molecular gas
in extreme ULIRGs \citep{Gracia06}. The results of this HCO$^+$ survey of
LIRGs and ULIRGs highlight the need of exploring the dense molecular
gas fraction in extreme ULIRGs like APM~08279+5255 with complementary
HCO$^+$ line observations. A comparison of HCO$^+$ and HCN line
intensities in high-z ULIRGs may allow the effects of excitation
to be disentangled from those of chemistry and help the derivation of
the physical conditions in these extraordinary objects.

Here we present the detection of HCO$^+$~(J=5--4) line emission in
APM~08279+5255 at z=3.911 made with the IRAM PdBI. This represents the
first detection of this molecular ion at such a high redshift. We
discuss in the context outlined above the implications of this result
for the interpretation of molecular line observations of gas-rich
high-redshift galaxies.



\section{Observations}
We observed APM~08279+5255 with the IRAM 6-element array in B
configuration on March 13, 2006. The spectral correlator was adjusted
to $z = 3.911$ and centered on redshifted HCO$^+$~(5--4) (rest
frequency at 445.903~GHz) to cover an effective bandwidth of 580 MHz
(equivalent to $\sim$1800km~s$^{-1}$). The size of the synthesized
beam was 1.46$\arcsec \times$1.20$\arcsec$ (PA=94$^\circ$) at the
observing frequency (90.797GHz).  APM~08279+5255 was observed for a
total integration time of 9.5 hrs on-source.  During the observations
the atmospheric phase stability on the most extended baselines was
always better than 20$^\circ$, typical for excellent winter
conditions. The absolute flux density scale was calibrated using
3C~84 and 0836+710, and should be accurate to better than
10\%. We have used the GILDAS package for the data reduction and
analysis.  The average 1$\sigma$--noise in channels of 20~MHz-width is
estimated to be 0.5~mJy~beam$^{-1}$. The phase tracking center of
these observations is at ($\alpha_o$, $\delta_o$)=(08$^{\rm h}31^{\rm
m}41.57^{\rm s}$, 52$^\circ$45$\arcmin$17.7$\arcsec$) coincident with
that used by Wagg et al.~(2005) in their HCN observations.  To
calculate luminosities, we assume a $\Lambda$--cosmology described by
H$_o$=70~km~s$^{-1}$, $\Omega_\Lambda$=0.7 and $\Omega_m$=0.3
\citep{Spergel03}. Throughout the paper, the velocity scale is
referred to the CO redshift z=3.911.

\section{Results}

\begin{figure}
\centering
\epsscale{1.0}
\plotone{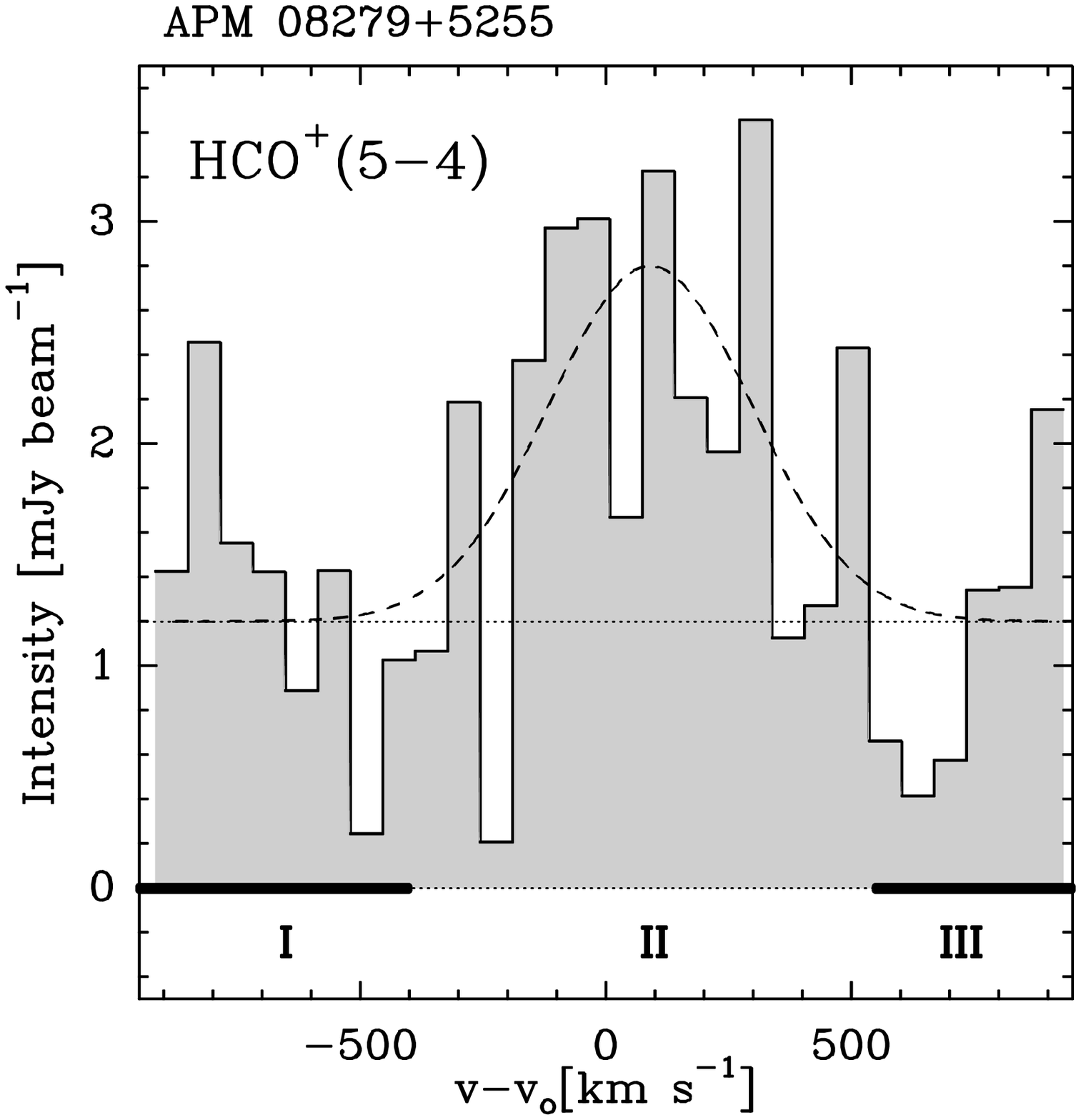}
\caption{Spectrum of the HCO$^{+}$~(5--4) line detected towards the peak of integrated line intensity in APM~08279+5255 at ($\alpha$, $\delta$)=(08$^{\rm h}31^{\rm m}41.73^{\rm s}$,  52$^\circ$45$\arcmin$17.4$\arcsec$). Velocities (v-v$_0$) have been re-scaled with respect to the CO redshift of z=3.911 \citep{Downes99}. The dashed line shows the Gaussian fit to the HCO$^{+}$~(5--4) emission and the highlighted channels in intervals I and III identify the range for line-free continuum emission.}
\label{spectrum}
\end{figure}

\begin{figure*}
\centering
\epsscale{1.0}
\plotone{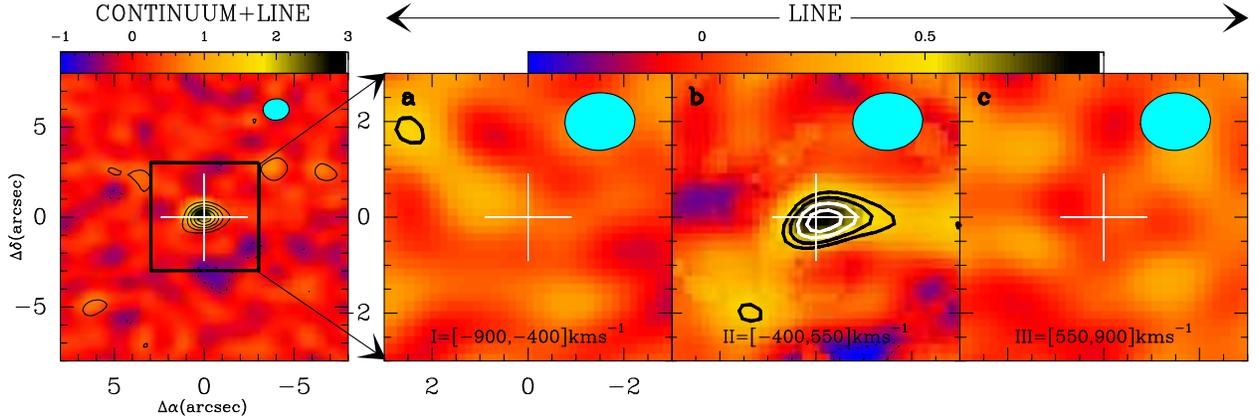}
\caption{The left panel shows the total (continuum+line) velocity-integrated emission map in APM~08279+5255. The {\it total} emission has been integrated from v-v$_0$=-900 to 900~km~s$^{-1}$. Levels are -3$\sigma$ (dotted), 3$\sigma$ to 15$\sigma$ in steps of 3$\sigma$ (1$\sigma$=0.17~Jy\,km~s$^{-1}$). Panels a--to--c show the HCO$^{+}$~(5--4) line emission maps obtained after subtraction of the continuum source for the three velocity intervals (I, II, III) defined in Figure~\ref{spectrum}: I=[-900,-400]~km~s$^{-1}$ (a), II=[-400,550]~km~s$^{-1}$ (b), III=[550,900]~km~s$^{-1}$ (c). Levels in panels a--to--c are -0.39 (dotted), 0.39, 0.52, 0.65, 0.78Jy\,km~s$^{-1}$ (equivalent to -3$\sigma$, 3$\sigma$, 4$\sigma$, 5$\sigma$, and 6$\sigma$ for channel II). To derive the continuum map we used channels I and III (see Figure~\ref{spectrum}). The filled ellipse in each panel represents the 1.46$\arcsec \times$1.20$\arcsec$ (PA=94$^\circ$) synthesized beam. ($\Delta\alpha$, $\Delta\delta$)--offsets in arcsec are relative to the peak of Continuum+Line emission at (08$^{\rm h}31^{\rm m}41.73^{\rm s}$,  52$^\circ$45$\arcmin$17.4$\arcsec$), identified by the thin cross.} 
\label{moments}
\end{figure*}

We have detected the HCO$^+$~(5--4) line and the dust continuum
emission (at 670$\micron$) both emitted in the
submillimeter range and redshifted to 3.3mm in APM~08279+5255.  The
overall results are summarized in Table~1. Figure~\ref{spectrum} shows
the spectrum of the HCO$^{+}$~(5--4) line (continuum not subtracted)
towards the peak of integrated intensity at ($\alpha$,
$\delta$)=(08$^{\rm h}31^{\rm m}41.73^{\rm s}$,
52$^\circ$45$\arcmin$17.4$\arcsec$) (see Figure~\ref{moments}). Within
the errors the peak position is in agreement with that determined from
previous mm-continuum maps \citep{Downes99, Wagg05}.  The velocity
coverage encompasses the full extent of the HCO$^+$ line emission.
Data from the line-free sidebands with velocities below
$-400$\,km~s$^{-1}$ and above $550$\,km~s$^{-1}$ (intervals I and III,
defined in Figure.~\ref{spectrum}), were combined with natural
weighting to estimate the continuum emission at 90.8~GHz to be
$\sim1.2\pm0.13$\,mJy, in good agreement with the value of
1.2$\pm$0.3\,mJy at 93.9~GHz of \citet{Downes99}.

The channel maps of Figure~\ref{moments} obtained for the three
velocity intervals defined in Figure~\ref{spectrum} (I, II and III)
clearly show that HCO$^+$~(5--4) line emission is only detected from
$-400$\,km~s$^{-1}$ to $550$\,km~s$^{-1}$ (interval II), with no
significant line emission outside this velocity range (i.e., in
intervals I or III).  The integrated HCO$^+$~(5--4) line intensity map
shows a $\sim$7$\sigma$ detection in APM~08279+5255 of
0.87$\pm$0.13\,Jy\,km~s$^{-1}$.  This translates into a line
luminosity
$L'_{HCO^+}$=$(3.5\pm0.6)\times\,10^{10}$\,K\,km~s$^{-1}$\,pc$^2$,
obtained from equation (3) of \citet{Solomon97}. The line emission
peaks roughly at the same position as the continuum. Within the
errors, HCO$^+$, HCN and high-J CO lines peak at the same
position. Furthermore, the line width (490$\pm$80~km~s$^{-1}$) and
central velocity (90$\pm$50~km~s$^{-1}$) of the HCO$^+$~(5--4) line
closely agree with those of the HCN line.  In particular, the central
velocities of the HCN and HCO$^+$ lines are both 80--90~km~s$^{-1}$
redshifted with respect to CO; this is an indication that HCN and
HCO$^+$ come from the same circumnuclear region in APM~08279+5255.

\section{Dense Gas in APM~08279+5255: excitation and inferred chemical abundances}

Taken at face value, the mere detection of CO~(4--3), CO~(9--8),  and mostly of
HCN~(5--4) and HCO$^{+}$~(5--4) line emission hints that there is a
significant reservoir of hot and very dense molecular gas in
APM~08279+5255. Partly implicit in this scenario, however, is the
assumption that the excitation of the HCN and HCO$^+$ lines is mainly
collisional. The inferred chemical abundances depend on this critical
assumption, as discussed below (Section~\ref{coll}). We study the
influence that non-collisional excitation of the J=5--4 lines of HCN
and HCO$^+$ may have on the derived chemical abundances for these
molecular species in Section~\ref{non-coll}.

\subsection{Collisional excitation}\label{coll}

\citet{Wagg05} performed escape-probability radiative transfer
calculations to re-examine the fit of \citet{Downes99} to the CO
9--8/4--3 ratio measured in APM~08279+5255 including also higher-J CO
transitions observed by Weiss et al. (2006, in preparation). Their
conclusion is that a single component model with a gas temperature
T$_{kin}\sim$80~K, a molecular hydrogen density
n(H$_2$)$\sim$4$\times$10$^{4}$~cm$^{-3}$ and a CO column density
N(CO)/$\Delta$v=4$\times$10$^{17}$~cm$^{-2}$ fits all the CO
ratios. The extrapolation of the CO model to match the observed
HCN~(5--4) line intensity leads Wagg et al. to conclude that the
[HCN]/[CO] abundance ratio in APM~08279+5255 (but not necessarily
the [HCN]/[H$_2$] ratio) is surprisingly large:
(1--2)$\times$10$^{-2}$. This is an order of magnitude larger than the
typical abundance ratio measured on small scales in galactic hot
cores. 

The detection of HCO$^+$~(5--4) line emission provides new constraints
on the chemistry of APM~08279+5255. In particular, we can estimate the
[HCN]/[HCO$^+$] abundance ratio in this source inside the collisional
excitation scheme with certain assumptions (the main one being the
adoption for HCO$^+$ and HCN of the same physical parameters derived
from CO lines).  The bottom line result issued from these calculations
is that the HCN~(5--4)/HCO$^+$~(5--4) luminosity ratio measured in
APM~08279+5255 ($\sim$1.1) is $\sim$3 times larger than expected if
the molecular abundances of the two species are assumed to be
comparable.  The critical densities of the HCO$^+$ v=0 ground
transitions are a factor of 10 lower than those of HCN. At equal
abundances and in the regime of subthermal excitation, HCO$^+$ v=0
J$>$1 rotational levels are expected to be more populated than those
of HCN. Therefore, if we aim at matching the line ratio reported
above, we need HCN to be significantly overabundant with respect to
HCO$^+$, by a factor of [HCN]/[HCO$^+$]$\sim$10.

We can explore the robustness of this conclusion by reconsidering some
of the underlying assumptions inherent in these calculations.  First,
we have implicitly assumed that both the filling factor and the
magnification factor of the HCO$^+$, HCN, and high-J CO lines are
similar. The similar line profiles and source parameters (size and
position) derived from HCO$^+$ and HCN are compatible with this
picture. It is conceivable, however, that the emission of HCO$^+$ and
HCN may come from gas with higher densities than derived from the
single-component CO model. In particular we have explored a range of
solutions matching the CO~(9--8)/CO~(4--3) ratio of APM~08279+5255
with higher densities and similar or slightly lower temperatures for
the gas.  An optimum fit can be found for
n(H$_2$)$\sim$4$\times$10$^{5}$~cm$^{-3}$ and T$_{kin}\sim$50-60~K. If
this solution is applied to HCN and HCO$^+$, in order to match the
HCN~(5--4)/HCO$^+$~(5--4) luminosity ratio of APM~08279+5255, we need
[HCN]/[HCO$^+$]$~\sim$10. In summary, the overabundance of
HCN with respect to HCO$^+$ is also required in the higher density
scenario.

\subsection{Non-collisional excitation: infrared pumping?}\label{non-coll}

The excitation of HCN and HCO$^+$ lines in a source with an infrared
luminosity as high as that of APM~08279+5255 may not be only
collisional, but also radiative. Like CO, the HCO$^+$ and HCN
millimeter line emissions are believed to arise in a thick
circumnuclear disk of radius 100-200 pc where a sizable fraction of
the dust is heated to $\simeq 200$ K by the AGN \citep{Lewis98,
Beelen06}.  The mid-IR emission from this disk ($\simeq 1$ Jy at 12
$\mu$m) is by far the brightest known in any known quasar to date. The
mid-IR radiation can excite the first $\nu_2=1$ bending modes of both
HCN ($\lambda= 14 \mu$m) and HCO$^+$ ($\lambda= 12 \mu$m) and, by
fluorescence, can populate the $J>3$ levels of the ground state. We
note that, prior to any IR pumping or collisional excitation, the
cosmic background temperature at the redshift of APM~08279+5255
(13.4~K) is high enough to populate the J=3 ground state level: the
fractional population of this level would be as large as 0.17 for a
rotational temperature $\sim$13.4~K. The radiative selection rules are
such that IR radiation will pump up molecules from this level to the
$\nu_2=1, J=4$ level, which will decay to the $\nu_2=0, J=5$
level. The frequencies and Einstein coefficients of the ro-vibrational
transitions of HCN and HCO$^+$ are similar ($\simeq$
1s$^{-1}$), so that both molecules behave similarly as concerns
radiative excitation.  Whether the J= 5 HCN and HCO$^+$ levels are
sufficiently populated at fluorescence equilibrium to explain the
observed lines depends then on the source geometry and on the
molecular abundances. \citet{Downes99} argue that the mid-IR source
must be smaller than the molecular source, because dust opacity
prevents the AGN radiation to penetrate far inside the molecular
torus. However, \citet{Nenkova02} argue that the disks are likely to
be clumpy and that the clumps throughout the disk are efficiently
heated by the AGN.  Such heating would explain the observed 200 K dust
component.  In addition, direct heating can occur within the torus, if
the latter hosts a strong starburst.

Assuming that the hot dust and the molecules are well mixed inside the
disk and that the dust is optically thick at $\sim$10$\mu$m, fluorescent
excitation could be quite effective. For a mid-IR source temperature
of $T_{d}$=200~K, the fraction of the sky subtended by this source at the
molecules needs only to be

$$f\geq \frac{A_{J=5-4}}{A_{v=1-0}} e^{-h\nu/kT_d} \simeq 0.1$$ 

for radiative pumping to be effective \citep{Carroll81}. 
It is interesting to note in this respect that the number of
12$\mu$m (and 14$\mu$m) photons escaping from the disk exceeds by a
factor of 5 the number of HCO$^+$ (HCN) photons observed in the J= 5-4
(670 $\mu$m) line. The opacity of the J$\geq$ 3, $\nu_2=0 \rightarrow
1$ ro-vibrational lines needs only to be $\sim 1$ to explain the
observed HCO$^+$ and HCN J= 5-4 line emissions, which for equal
molecular abundances will have similar intensities. Such an opacity
will be reached for column densities of HCO$^+$ and HCN of N(X)/$\Delta v
\simeq$10$^{13}$ cm$^{-2}$, or $10^{-4}$ times the CO column density
derived by \citet{Wagg05}. In summary, inside the infrared pumping scheme,
we do not need the abundance of HCN to be anomalously high with respect to CO
and HCO$^+$.

\section{Discussion and Conclusions}

The two scenarios invoked above, both accounting for the {\it
unexpectedly} large HCN(5--4)/HCO$^+$(5--4) ratio measured in
APM~08279+5255, have completely different but equally relevant
implications for the interpretation of high-J molecular line
observations of dense gas in other high-redshift galaxies.

Inside the collisional excitation model it is implied that the
abundance of HCN in APM~08279+5255 is {\it anomalously} high with
respect to CO and HCO$^+$.  Different mechanisms can increase the 
abundance of HCN with respect to CO and HCO$^+$ \citep[e.g., see
discussion of][]{Gracia06}. This includes high-ionization chemistry
driven by X-rays around an AGN, chemical enhancement of HCN in 
star-forming regions  and, if molecules coexist with the hot dust, 
enhanced destruction of HCO$^+$ by reaction with water wapor evaporated from
the dust. The ingredients (an AGN, a massive star forming episode 
and hot dust) are  all present in APM~08279+5255,
even though X-ray ionization seems unlikely to raise [HCN]/[HCO$^+$] 
well above 1 \citep{Lepp96}. We can also tentatively discard selective depletion
of oxygen bearing species as an explanation for the inferred
[HCN]/[HCO$^+$] abundance ratio in APM~08279+5255. \citet{Usero04}
analyzed the case of the Seyfert 2 galaxy NGC~1068 and found that the
solution provided by oxygen depletion models matching an abundance
ratio of [HCN]/[CO]$\sim$10$^{-3}$ (i.e., $\sim$10 times lower than that
of APM~08279+5255) leads to an excessively large [HCN]/[HCO$^+$]
abundance ratio ($>$20).

Of particular note, the advantage of the radiative excitation scheme
over the collisional excitation model is double. First, radiative
excitation does not require very high gas densities to explain the
HCN/HCO$^+$ line ratios observed in APM~08279+5255, even through 
the CO line ratios seem to imply medium-high densities. Furthermore,
infrared pumping can also account for the near equality of the HCN and
HCO$^+$ J=5-4 line intensities while keeping an abundance
ratio [HCN]/[HCO$^+$]  within the range 0.5-2, typical of that observed in
dense molecular clouds in the Galaxy and in nearby galaxies
\citep[e.g., see Table~8 of][]{Wang04}. 

The availability of only one observed transition of HCN and
HCO$^+$ prevents us from making a more educated choice of the right
scenario for APM~08279+5255. Future observations that include lower
and higher-J transitions of both HCO$^+$ and HCN will be required to
confront a set of observed line ratios with the predictions issued
from chemical and excitation models.


\acknowledgments SGB thanks the support from the Spanish
MEC and Feder funds under grant ESP2003-04957 and SEPCT/MEC under
grant AYA2003-07584. We thank W. Klemperer, E. Herbst, A. Fuente and M. Elitzur for
comments on our work. 



\clearpage

\clearpage

\clearpage

\begin{deluxetable}{lc}
\tablecaption{Observational results}
\tablewidth{0pt}
\tablecolumns{2}
\tablehead{\colhead{Parameter} & \colhead{Value}}
\startdata
$\alpha_{J2000}$ & 08$^{\rm h}31^{\rm m}41.73^{\rm s}\pm0.01^{\rm s}$ \\
$\delta_{J2000}$ & 52$^\circ45'17.4''\pm0.1''$\\
$V_{HCO^+(5-4)}$\tablenotemark{a} & $90\pm50$ \,km~s$^{-1}$ \\
$I_{HCO^+(5-4)}$& 0.87$\pm$0.13\,Jy\,km~s$^{-1}$ \\
$L'_{HCO^+(5-4)}$\tablenotemark{b} & $(3.5\pm0.6)\times\,10^{10}$\,K\,km~s$^{-1}$\,pc$^2$\\
90.8GHz continuum & 1.2$\pm$0.13\,mJy 
\enddata
\tablenotetext{a}{Velocity referred to z=3.911 (derived from CO lines)}
\tablenotetext{b}{Uncorrected for lensing}
\end{deluxetable}


\begin{thebibliography}{}

\bibitem[Beelen et al.(2006)]{Beelen06} Beelen, A., Cox, P., 
Benford, D.~J., Dowell, C.~D., Kovacs, A., Bertoldi, F., Omont, A., \& 
Carilli, C.~L.\ 2006, ArXiv Astrophysics e-prints, arXiv:astro-ph/0603121 


\bibitem[Carroll \& Goldsmith(1981)]{Carroll81} Carroll, T.~J., 
\& Goldsmith, P.~F.\ 1981, \apj, 245, 891 

\bibitem[Downes et al.(1999)]{Downes99} Downes, D., Neri, R., 
Wiklind, T., Wilner, D.~J., \& Shaver, P.~A.\ 1999, \apjl, 513, L1

\bibitem[Gao \& Solomon(2004a)]{Gao04a} Gao, Y., \& Solomon, 
P.~M.\ 2004a, \apjs, 152, 63 

\bibitem[Gao \& Solomon(2004b)]{Gao04b} Gao, Y., \& Solomon, 
P.~M.\ 2004b, \apj, 606, 271 

\bibitem[Graci{\'a}-Carpio et al.(2006)]{Gracia06} 
Graci{\'a}-Carpio, J., Garc{\'{\i}}a-Burillo, S., Planesas, P., \& Colina, 
L.\ 2006, \apjl, 640, L135 

\bibitem[Lepp \& Dalgarno(1996)]{Lepp96} Lepp, S., \& 
Dalgarno, A.\ 1996, \aap, 306, L21 
 
\bibitem[Lewis et al.(1998)]{Lewis98} Lewis, G.~F., Chapman, 
S.~C., Ibata, R.~A., Irwin, M.~J., \& Totten, E.~J.\ 1998, \apjl, 505, L1 

\bibitem[Lewis et al.(2002)]{Lewis02} Lewis, G.~F., Carilli, 
C., Papadopoulos, P., \& Ivison, R.~J.\ 2002, \mnras, 330, L15 

\bibitem[Maloney et al.(1996)]{Maloney96} Maloney, P.~R., 
Hollenbach, D.~J., \& Tielens, A.~G.~G.~M.\ 1996, \apj, 466, 561 
 
\bibitem[Nenkova et al.(2002)]{Nenkova02} Nenkova, M., 
Ivezi{\'c}, {\v Z}., \& Elitzur, M.\ 2002, \apjl, 570, L9 

\bibitem[Papadopoulos et al.(2001)]{Papadopoulos01} Papadopoulos, P., 
Ivison, R., Carilli, C., \& Lewis, G.\ 2001, \nat, 409, 58 

\bibitem[Solomon et al.(1992)]{Solomon92} Solomon, P.~M., Downes, 
D., \& Radford, S.~J.~E.\ 1992, \apjl, 387, L55 
 
\bibitem[Solomon et al.(1997)]{Solomon97} Solomon, P.~M., Downes, 
D., Radford, S.~J.~E., \& Barrett, J.~W.\ 1997, \apj, 478, 144 

\bibitem[Spergel et al.(2003)]{Spergel03} Spergel, D.~N., et al.\ 
2003, \apjs, 148, 175 
 
\bibitem[Sternberg et al.(1994)]{Sternberg94} Sternberg, A., 
Genzel, R., \& Tacconi, L.\ 1994, \apjl, 436, L131 

\bibitem[Usero et al.(2004)]{Usero04} Usero, A., 
Garc{\'{\i}}a-Burillo, S., Fuente, A., Mart{\'{\i}}n-Pintado, J., \& 
Rodr{\'{\i}}guez-Fern{\'a}ndez, N.~J.\ 2004, \aap, 419, 897 

\bibitem[Wagg et al.(2005)]{Wagg05} Wagg, J., Wilner, D.~J., 
Neri, R., Downes, D., \& Wiklind, T.\ 2005, \apjl, 634, L13 

\bibitem[Wang et al.(2004)]{Wang04} Wang, M., Henkel, C., 
Chin, Y.-N., Whiteoak, J.~B., Hunt Cunningham, M., Mauersberger, R., \& 
Muders, D.\ 2004, \aap, 422, 883 

\bibitem[Wu et al.(2005)]{Wu05} Wu, J., Evans, N.~J., Gao, 
Y., Solomon, P.~M., Shirley, Y.~L., \& Vanden Bout, P.~A.\ 2005, \apjl, 
635, L173 
 

\end{thebibliography}
\end{document}